\tolerance   10000
\magnification = 1200
\baselineskip=1.65\normalbaselineskip
\font\ti=cmti10  scaled 1000

\font\bigrmtwenty = cmb10 scaled 1500 
\font\bigrmsixteen = cmb10 scaled 1000
\font\smallrm = cmr10 scaled 800
\font\smallrmb = cmb10 scaled 800
\font\name = cmr10 scaled 1200
\input epsf
\let\origeqno=\eqno
\def\eqno(#1){\origeqno (\rm #1)}

\noindent
{\centerline {\bigrmtwenty PARTIAL AVERAGING AND RESONANCE TRAPPING  }}
\hfill
\noindent 
{\centerline {\bigrmtwenty  IN A RESTRICTED THREE-BODY SYSTEM }}
\vskip   25pt
\noindent
{\centerline  {\bf{\name NADER HAGHIGHIPOUR }}}
\hfill
\noindent
{\centerline{\ti {Department of Physics and Astronomy, 
             Northwestern University, }}}
\hfill
\vskip -25pt
\noindent
{\centerline {\ti { Evanston, Illinois$\>$ 60208 , U.S.A.}}}
\vskip  20pt
\noindent
{\bigrmsixteen  Abstract.}
\vskip 15pt
\vbox{
\baselineskip=1.55\normalbaselineskip

Based on the value of the orbital eccentricity of a particle and also its
proximity to the exact resonant orbit in a three-body system, the Pendulum
Approximation (Dermott $\&$ Murray 1983) or the Second Fundamental
Model of Resonance (Andoyer 1903; Henrard $\&$ Lema\^\i tre 1983)
are commonly used to study the motion of that particle near its resonance
state. In this paper, we present the method of  
partial averaging as an analytical approach to study the dynamical
evolution of a body near a resonance.
To focus attention on the capabilities of this technique,
a restricted, circular
and planar three-body system is considered and the dynamics of its outer planet
while captured in a resonance with the inner body is studied. 
It is shown that the first-order
partially averaged system resembles a mathematical pendulum whose 
librational motion can be viewed as a geometrical interpretation
of the resonance capture phenomenon. The driving force of this pendulum 
corresponds to the gravitational attraction of the inner body and its
contribution, at different resonant states,  
is shown to be proportional
to $e^s$, where $s$ is the order of the resonance
and $e$ is the orbital eccentricity of the outer planet. As examples of
such systems, the cases of (1:1), (1:2) and (1:3) resonances are discussed and
the results are compared with known planetary systems
such as the Sun-Jupiter-Trojan asteroids.}
\vskip 10pt
\noindent
{\bigrmsixteen Key words:} celestial mechanics, planetary dynamics, 
resonance capture, averaging.
\vskip 16pt
\noindent
{Email Address: nader@northwestern.edu}

\vfill
\eject

\noindent
{\bigrmsixteen {1 $\>$ INTRODUCTION}}
\vskip  5pt

The study of the dynamical evolution of a planetary system 
while captured in a resonance has a long history in dynamical astronomy.
Since the pioneering work of Poincar\'e (1902) on the study of the 
near-resonance motions in a restricted three-body system by means of a
zeroth-order resonance Hamiltonian
(i.e., a Hamiltonian with no perturbing terms other than the resonant
ones), the body of literature produced on this subject 
has become so rich and extensive that it is virtually impossible to cite
all the articles here. Recent discoveries of extrasolar planetary systems 
such as  Gliese 876 (Marcy et al. 2001), where
two planets are locked in a near (2:1) commensurability, 
have also provided rich grounds for astrodynamicists to extend such 
studies to the systems beyond the boundaries of our solar system 
(Laughlin $\&$ Chambers 2001; Lissauer $\&$ Rivera 2001;
Murray, Paskowitz $\&$ Holman 2001; Snellgrove, Papaloizou $\&$ Nelson 2001;
Lee $\&$ Peale 2001). 

There are two analytical approaches that are commonly
taken in the study of the dynamics of the bodies of a three-body 
system while captured in a resonance. For a particle 
with high orbital eccentricity $(\geq 0.15)$ in 
a first-order resonance, or for a particle at any resonance
with higher orders, the Pendulum Model
is used when the particle's
orbit is sufficiently close to the real resonant location
(Dermott $\&$ Murray 1983). The Hamiltonian Model, or as it is often called,
the Second Fundamental Model of Resonance (Andoyer 1903; Henrard $\&$
Lema\^\i tre 1983) is usually used for small eccentricities.
A comparison of the results of the application of these two models
to the study of the motion of a test particle in a
first-order interior resonance as well as an
exterior $(1:n'), n'=2,3,4,5$ commensurability,  
with the results of numerical integrations can be found in a series
of papers by Winter $\&$ Murray (1997a$\&$b). In these articles,
as a part of their comprehensive project CRISS-CROSS on understanding
the location and origin of chaotic regions in the phase space of
our solar system, Winter and Murray present a detailed analysis of 
the dynamics of a test particle at resonance.

The purpose of this paper is to present a relatively new approach, 
namely the method of partial averaging near a resonance,
to study analytically the dynamics of a system near a resonant
state. This technique that is based on the Averaging Theorem 
(see Sanders $\&$ Verhulst 1985 and also Wiggins 1996),
enables one to avoid certain complexities 
by studying the behavior of the 
system averaged  over a fast angular variable (Melnikov 1963; 
Guckenheimer $\&$ Holmes 1983; Greenspan $\&$ Holmes 1983). 
It is necessary to mention that
a complete picture of the dynamical evolution of the system
can only be obtained by direct analysis of its equations of motion. 
The partially averaged system allows one to 
focus attention on the slow-changing quantities.
Such an idea is commonly used in celestial mechanics:
the Hamiltonian of the system is averaged over a fast variable and 
the resulting averaged Hamiltonian is used to study the slow dynamics
of the system. A review of this technique can be found in the work of Ferraz-Mello
(1997) and the references cited therein. To demonstrate the capabilities of 
the method of partial averaging, a restricted, 
circular and planar three-body system is considered here and the
motion of its outer planet near an $(n:n')$ resonance is studied.

Although the method of partial averaging near a resonance
has long been used 
by mathematicians in their studies of dynamical 
systems near resonances
(Melnikov 1963; Guckenheimer $\&$ Holmes 1983; Greenspan $\&$ Holmes 1983), 
its application to astronomical systems is quite
recent. One can find such applications in papers by this author on
study of the dynamical evolution of a planetary system in a uniform and
homogeneous disk of planetesimals (Haghighipour 1999, 2000)
and also in a series of articles by 
Chicone, Mashhoon $\&$ Retzloff (1996a$\&$b, 1997a$\&$b, 1999, 2000) 
on their extensive study of 
the dynamics of a binary system subject to incident gravitational radiation 
as well as gravitational radiation damping. 

The system of interest in this paper is a 
hypothetical restricted, planar and circular three-body
system consisting of a star and two planets. 
The equations of motion of the outer planet of this system
are presented in section 2. Section 3 deals with the system at
resonance. In section 4, the method of partial averaging near a
resonance is
applied and the averaged dynamics of the outer planet, in
the first order of perturbation, is studied. It is shown in
this section that the contribution of the gravitational attraction 
of the inner planet on the averaged dynamics of the outer body 
at different resonant states is directly proportional to
the orbital eccentricity of the latter
with a power equal to the order of
the resonance. As examples of such cases, 
the (1:1), (1:2) and (1:3) resonances are studied in detail
and a comparison with the system of Sun-Jupiter-Trojan asteroids
as an actual case of a near (1:1) commensurability is presented.
Section 5 concludes this study by reviewing the results and 
presenting remarks on their applicability to other planetary systems.
\vskip  25pt
\noindent
{\bigrmsixteen {2 $\>$ THE SYSTEM}}
\vskip  5pt 

The system of interest in this study is a planar three-body system
consisting of a star $S$ and two planets $P_1$ and $P_2$ where 
$P_2$ is the outer planet 
and $P_1$, the more massive planet, orbits $S$ uniformly along a 
circular path. 
The orbital period of ${P_1}$ is considered to be known and constant.
It is also assumed that the  mass of $S$ is so much larger than $P_1$ and $P_2$ that 
the effect of their gravitational attraction on $S$ can be neglected. 

As mentioned earlier, it is
the dynamics of $P_2$ that is of interest here.
In an inertial coordinate system with its origin at $S$ and its axes on 
the plane of the system,
the equation of motion of the outer planet can be written as
\vskip  2pt
$$
{m_2}\,{{{d^2}{{\vec r}_2}}\over {d{t^2}}}\,+\,
{\cal G}\,{{M {m_2}}\over {|{{\vec r}_2}|^3}}\,{{\vec r}_2}\,+\,
{\cal G}\, {{{m_1} {m_2}}\over {|{{\vec r}_2} - {{\vec r}_1}|^3}} ({{\vec r}_2} 
- {{\vec r}_1})\,=\,0\>\>.
\eqno  (1)
$$
\vskip  5pt
\noindent
In this equation,  $\cal G$ is the gravitational constant,
${\vec r}_1$ and ${\vec r}_2$ are the position vectors of the two planets, 
$m_1$ and $m_2$ are their corresponding masses and $M$ is the mass of the 
central star. 
For future purposes, it is more convenient to write equation (1) 
in a dimensionless form. 
Introducing $r_0$ and $t_0$ as the quantities that carry units of 
length and time, respectively, 
equation (1) can be written as
\vskip  2pt
$$
{{{d^2}{\vec {\hat r}}}\over {d{{\hat t}^2}}}\,+\,
{\hat k}\,{{\vec {\hat r}}\over {|{\vec {\hat r}}|^3}}\,+\,
\mu\,{\hat k}\,{{({\vec {\hat r}}-{{\vec {\hat r}}_1})}\over
{|{\vec {\hat r}} - {{\vec {\hat r}}_1}|^3}}\,=\,0 \>\>\>,
\eqno  (2)
$$
\vskip  5pt
\noindent
where ${\vec {\hat r}}\,,{{\vec{\hat r}}_1}$ and $\hat t$ are dimensionless
quantities given by ${{\vec r}_2}\,=\,{r_0}\,{\vec {\hat r}}\,,
{{\vec r}_1}\,=\,{r_0}\,{{\vec {\hat r}}_1}$ and 
$t\,=\,{t_0}\,{\hat t}$. In this 
equation, ${\hat k}={\cal G}M{t_0^2}/{r_0^3}$ and $\mu={m_1}/M$. 
The assumption of a
uniform circular motion for $P_1$ allows one to set ${\hat k}=1$
by choosing ${r_0}={r_1}$ and ${t_0}={T_1}/2\pi$ where $T_1$
is the orbital period of $P_1$.

As mentioned in the previous section, we would like to analyze
equation (2) using the method of partial averaging. To do so,
it is necessary to write equation (2) in terms of appropriate  
action-angle variables.
The most appropriate action-angle variables for this purpose are
the Delaunay variables given by 
$L = {a^{1/2}}\,,\,
G = {{\cal P}_\theta}= {[a(1-{e^2})]^{1/2}}\,,\,\ell = u-e \sin u$ 
and $g = \theta - v$ where $e={{\cal P}_r}{{\cal P}_\theta}/\sin v$ and $a$ 
are the eccentricity and the semimajor axis of the 
osculating ellipse of $P_2$, $\theta$ is its plane-polar angle, 
${\cal P}_r$ and ${\cal P}_\theta$ are its radial and angular momenta
and $v$ and $u$
are its eccentric and true anomalies related as
$$
r\,=\,{{G^2}\over {1+e \cos v}}\,=\,a\,(1-e \cos u)\>.
\eqno  (3)
$$
\vskip  6pt
\noindent
In order to write equation (2) in term of the Delaunay variables,
it is more convenient to first write this equation in terms of  
${{\cal P}_r}$ and ${{\cal P}_\theta}$. That is,
\vskip  2pt
$$\!\!\!\!\!\!\!\!\!\!\!\!\!\!\!\!\!\!\!\!\!\!\!\!\!\!\!\!\!\!\!\!\!\!\!\!
\!\!\!\!\!\!\!\!\!\!\!\!\!\!\!\!\!\!\!\!\!\!\!\!\!\!\!\!\!\!\!\!\!\!\!\!\!\!\!\!
\!\!\!\!\!\!\!\!\!\!\!\!\!\!\!\!\!\!\!\!\!\!\!\!\!\!\!\!\!\!\!\!\!\!\!\!\!
{{\cal P}_r}\,=\,{\dot r}\>,
\eqno  (4)
$$
$$\!\!\!\!\!\!\!\!\!\!\!\!\!\!\!\!\!\!\!\!\!\!\!\!\!\!\!\!\!\!\!\!
\!\!\!\!\!\!\!\!\!\!\!\!\!\!\!\!\!\!\!\!\!\!\!\!\!\!\!\!\!\!\!\!\!\!\!\!\!\!
{\!\!\!\!\!\!\!\!\!\!\!\!\!\!\!\!\!\!\!\!\!\!\!\!\!\!\!\!\!\!\!\!\!\!\!\!
{\cal P}_\theta}\,=\, {r^2}\, {\dot \theta} \>,
\eqno  (5)
$$
$$
{{\dot {\cal P}}_r}\,=\,{1\over {r^3}}\,{{\cal P}_\theta^2}\,-\,{1\over {r^2}}\,-\,
{\mu\over{|{\vec r}\,-\,{\bf{{\vec r}_1}}|^3}}\>
\Bigl[{\vec r}\,-\,\cos (\theta\,-\,{\theta_1})\Bigr]\>,
\eqno (6)
$$
$$\!\!\!\!\!\!\!\!\!\!\!\!\!\!\!\!\!\!\!\!\!\!\!\!\!\!\!\!\!\!\!\!\!\!\!\!\!\!\!\!
\!\!\!\!\!\!\!\!\!
{{\dot {\cal P}}_\theta}\,=\,-\,{\mu\over{|{\vec r}\,-\,{\bf {{\vec r}_1}}|^3}}\,
r\,\sin (\theta\,-\,{\theta_1})\>,
\eqno  (7)
$$
\vskip  10pt
\noindent
where ${\theta_1}={\omega_1}{\hat t}$ and ${\omega_1}=1$, 
is the dimensionless angular velocity of $P_1$. 
In equations (4) to (7), the hat signs have been
dropped for the sake of simplicity and the overdot indicates a derivative
with respect to the dimensionless time $\hat t$. The vector
$\bf{{\vec r}_1}$ in equations (6) and (7) is the unit vector along ${\vec r}_1$. 
In terms of the Delaunay variables, equations (4) to (7) can be written as
\vskip 2pt
$$\!\!\!\!\!\!\!\!\!\!\!\!\!\!\!\!\!\!\!\!\!\!\!\!\!\!\!\!\!\!\!\!\!\!\!\!\!\!\!\!
\!\!\!\!\!\!\!\!\!\!\!\!\!\!\!\!\!\!\!\!\!\!\!\!\!\!\!\!\!\!\!\!\!\!\!\!\!\!\!\!
\!\!\!\!\!\!\!\!\!\!\!\!\!\!\!\!\!\!\!\!\!\!\!\!\!\!\!\!\!\!\!\!\!\!\!\!\!\!\!\!
\!\!\!\!\!\!\!\!\!\!\!\!\!\!\!\!\!\!\!\!\!\!\!\!\!\!\!\!\!\!\!\!\!\!\!\!\!\!\!\!
\!\!\!\!\!\!\!\!\!\!\!\!\!\!\!\!\!\!\!\!\!\!\!\!\!\!\!\!\!\!\!\!\!\!\!\!
{\dot G}\,=\,r\,{F_\theta}\>\>\>,
\eqno  (8)
$$
$$\!\!\!\!\!\!\!\!\!\!\!\!\!\!\!\!\!\!\!\!\!\!\!\!\!\!\!\!\!\!\!\!\!\!\!\!\!\!\!\!
\!\!\!\!\!\!\!\!\!\!\!\!\!\!\!\!\!\!\!\!\!\!\!\!\!\!\!\!\!\!\!\!\!\!\!\!\!\!\!\!\!\!
\!\!\!\!\!\!\!\!\!
{\dot L}\,=\,a\,(1-e^2)^{-{1/ 2}}\>\Bigl[{F_\theta}\,+\,
e\,({F_r}\,\sin v\,+\,{F_\theta}\,\cos v)\Bigl]\>\>\>,
\eqno  (9)
$$
$$\!\!\!\!\!\!\!\!\!\!\!\!\!\!\!\!\!\!\!\!\!\!\!\!\!\!\!\!\!\!\!\!\!\!\!\!\!\!\!\!
\!\!\!\!\!\!\!\!\!\!\!\!\!
{\dot g}\,=\,{1\over e}\,\bigl[a(1-e^2)\bigr]^{1/ 2}\,
\biggl[\,{F_\theta}\Bigl({{\sin v}\over {1+e \cos v}}\Bigr)\,-\,
({F_r}\,\cos v\,-\,{F_\theta}\,\sin v)\biggl]\>\>\>,
\eqno  (10)
$$
$$
{\dot \ell}\,=\,{a^{-3/2}}\,+\,{r\over e}\,{a^{-{1/ 2}}}\,
\biggl\{({F_r}\,\cos v\,-\,2\,{F_\theta}\,\sin v)\,+\,
{1\over 2}\,e\,\Bigl[({F_r}\,\cos 2 v \,-\,{F_\theta}\,\sin 2 v)\,-\,
3\,{F_r}\Bigr]\biggr\}\>\>\>,
\eqno (11)
$$
\vskip  5pt
\noindent
where 
\vskip  2pt
$$
{F_r}\>=\,-\,\mu\>{{r\,-\,\cos(\theta-{\theta_1})}\over {|{\vec r}-
{\bf{\vec r_1}}|^3}}\qquad,\qquad
{F_\theta}\>=\>-\,\mu\,{{\sin(\theta-{\theta_1})}\over {|{{\vec r}-
{\bf{\vec r_1}}|^3}}}\>\>\>.
\eqno  (12)
$$ 
\vskip  40pt
\noindent
{\bigrmsixteen {3 $\>$ SYSTEM AT RESONANCE}}
\vskip  10pt

Consider an $(n:n')$ commensurability between 
the angular frequency of the inner planet $\omega_1$, and
${\omega_\ell}={a^{-3/2}}$, the Keplerian frequency of the 
osculating ellipse of the outer one.  That is,
$$
n\,{\omega_1}\,=\,{n'}\,{\omega_\ell}\>.
\eqno (13)
$$
\noindent
For our restricted circular system with this resonance condition,
the Keplerian frequency $\omega_\ell$, and therefore, the
semimajor axis of the outer planet at resonance are constant. 
The constant value of the semimajor axis, denoted by
$a_{(n:n')}$, corresponds to
the resonant value of the action variable $L$ as
${L_{(n:n')}}={a_{(n:n')}^{1/2}}$. We would like to study
the dynamics of the outer planet when $L$ varies in the vicinity of
this value. For this purpose, we need to introduce 
an appropriate transformation that renders equations
(8) to (11) in a form that includes deviations of $L$ from 
$L_{(n:n')}$. Let $D$ be an action variable  measuring
these deviations. We then write,
$$
L\,=\,{a_{(n:n')}^{1/2}}\,+\,{\mu^{1/2}}\,D\,,
\eqno  (14)
$$
\noindent
and
$$
\ell\,=\,{a_{(n:n')}^{-3/2}}\,{\hat t}\,+\,\varphi\,,
\eqno (15)
$$
\noindent
where $\varphi$ denotes 
the deviations of the mean anomaly $\ell$ from its Keplerian
value (Appendix A). Such transformations have been repeatedly
used in application of Hamiltonian averaging techniques
to resonant systems (Ferraz-Mello 1997).
It is important to mention that the choice
of $\mu^{1/2}$ in equation (14) is to assure equal lowest 
order of perturbation for
$\dot D$ and $\dot \varphi$ after writing
equations (8) to (11) for the system near resonance. 
Details on this can be found in  Wiggins (1996)
and also in Haghighipour (2000).

For the purpose of writing equations (8) to (11) near
a resonance, it is more convenient to write these equations
as (Appendix B)
$$
{\dot L}\,=\,-\,\mu\,{{\partial H}\over {\partial \ell}}\>,
\qquad\qquad\qquad\qquad
{\dot G}\,=\,-\,\mu\,{{\partial H}\over {\partial g}}\>,
\eqno  (16)
$$
\noindent
and
$$\!\!\!\!\!\!\!\!\!\!\!\!\!
{\dot \ell}\,=\,{\omega_\ell}\,+\,\mu\,{{\partial H}\over {\partial L}}\>,
\qquad\qquad\qquad\qquad
{\dot g}\,=\,\mu\,{{\partial H}\over {\partial G}}\>,
\eqno  (17)
$$
\vskip  2pt
\noindent
where $H=-|{\vec r}-{\bf{{\vec r}_1}}|^{-1}$
(hereafter, {\it external Hamiltonian}) 
represents the perturbative effect of the inner planet, $P_1$.
The dynamical equations of the system near resonance
can now be written as (Haghighipour 1999)
\vskip 2pt
$$\!\!\!\!\!\!\!\!\!\!\!\!\!\!\!\!\!\!\!\!\!\!\!\!\!\!\!\!\!\!
\!\!\!\!\!\!\!\!\!\!\!
{\dot {D}}\,=\,-\,{\mu^{1/2}}\,
{{\partial H}\over {\partial \ell}}\,-\,
\mu\,D\,{{{\partial^2}H}\over {\partial \ell \partial L}}\,+\,
O({\mu^{3/2}})\>,
\eqno  (18)
$$
$$\>\>\>\>\>
{\dot {\varphi}}\,=\,-\,3\,{\mu^{1/2}}\,{a_{(n:n')}^{-2}}\,D\,+\,
\mu\,\Bigl[6{a_{(n:n')}^{-5/2}}\,{D^2}\,+\,{{\partial H}\over {\partial L}}\Bigr]
\,+\,O({\mu^{3/2}})\>,
\eqno (19)
$$
$$\!\!\!\!\!\!\!\!\!\!\!\!\!\!\!\!\!\!\!\!\!\!\!\!\!\!\!\!\!\!
\!\!\!\!\!\!\!\!\!\!\!\!\!\!\!\!\!\!\!\!\!\!\!\!
\!\!\!\!\!\!\!\!\!\!\!\!\!\!\!\!\!\!\!\!\!\!\!\!\!\!\!\!\!\!\!
{\dot {G}}\,=\,-\,\mu\,{{\partial H}\over {\partial g}}\,+\,
O({\mu^{3/2}})\>,
\eqno (20)
$$
$$\!\!\!\!\!\!\!\!\!\!\!\!\!\!\!\!\!\!\!\!\!\!\!\!\!\!\!\!\!\!
\!\!\!\!\!\!\!\!\!\!\!\!\!\!\!\!\!\!\!\!\!\!\!\!\!\!\!\!\!\!
\!\!\!\!\!\!\!\!\!\!\!\!\!\!\!\!\!\!\!\!\!\!\!\!\!\!\!\!\!\!\!\!
{\dot {g}}\,=\,\mu\,{{\partial H}\over {\partial G}}\,+\,
O({\mu^{3/2}})\>.
\eqno (21)
$$
\vskip  3pt
\noindent
In these equations, all partial derivatives are evaluated at
$({L_{(n:n')}},G,{L_{(n:n')}^{-3}}{\hat t}+\varphi,g)$.
The averaged dynamics of the system is obtained by applying 
the averaging technique presented in appendix A to equations 
(18) to (21). 
\vskip  25pt
\noindent
{\bigrmsixteen {4 $\>$ FIRST-ORDER AVERAGED SYSTEM}}
\vskip  5pt 

As mentioned before, we would like to study the averaged
dynamics of the outer planet near a resonance, in the
first order of the perturbation parameter $\mu^{1/2}$.
In that order, equations (18) to (21) are written as
\vskip 1pt
$$
{\dot {D}}\,=\,-\,{\mu^{1/2}}\,{\cal F}\,,\qquad\qquad
{\dot {\varphi}}\,=\,-\,{\mu^{1/2}}\,\Bigl[{{3D}\over 
{a_{(n:n')}^2}}\Bigr]\,,\qquad\qquad
{\dot {G}}\,=\,{\dot {g}}\,=\,0\>,
\eqno  (22)
$$
\vskip  5pt
\noindent
where ${\cal F}\,=\,\partial H/\partial \ell$,  is evaluated at
$({L_{(n:n')}},G,{L_{(n:n')}^{-3}}{\hat t}+\varphi,g)$. 

In this section, we apply the method of partial averaging, as described
in appendix A, to equations (22). As mentioned in that appendix,
the analysis presented there is only valid for systems with one angular 
variable. An inspection of the main dynamical equations of the outer
planet (i.e., equations (16) and (17)) reveals that 
these equations along with ${{\dot \theta}_1}= 1$, represent the
time variations of two action variables $L$ and $G$ and three
angular variables $\ell, g$ and $\theta_1$. As shown by equations
(22), to the first order of $\mu^{1/2}\,,\, {\dot g}=0$. Also,  
from the resonance condition (13) and the transformation (15), 
the angular variables $\ell$ and ${\theta_1}={\hat t}$ 
are related as 
$\ell\,=\,(n/n'){\hat t}+\varphi$. This relation implies that
equations (22) represent a dynamical system with an action
variable $D({\hat t})$ and an angular variable $\varphi({\hat t})$.
These equations are now in the correct form for applying the
partial averaging technique. 
Using formula (A7), the averaged dynamics of the
outer planet, to the first order of perturbation, can be written as
$$
{\ddot {\bar \varphi}}\,-\,3\,\mu\,{a_{(n:n')}^{-2}}
{\bar{\cal F}}({L_{(n:n')}}\,,\,G\,,\,{\bar \varphi}\,,\,g)\,=\,0\>,
\eqno  (23)
$$
\noindent
where the overbar denotes an averaged quantity.

To study this equation, it only remains to calculate $\cal F$ which
requires one to express $H$ in terms of the mean 
anomaly $\ell$. From its definition, $H$ can be written as
\vskip  2pt
$$
H\,=\,-\,\Bigl[1\,+\,{r^2}\,-\,2\,r\,\cos (\theta\,-\,{\theta_1})
\Bigr]^{-1/2}\>\>.
\eqno  (24)
$$
\vskip 2pt
\noindent
Substituting for $r$ from equation (3) and replacing $\theta$ by $g+v$,
we have
$$\eqalign {
H\,=\,-\,\Biggl\{
&\biggl[1+{a^2}-2a \cos (\ell + g - {\theta_1})\biggr]\cr
&-\,e\,a\,\biggl[2 a \cos \ell +  \cos (2\ell+g-{\theta_1})-
3 \cos (g-{\theta_1})\biggr]\cr
&+\,{1\over 4}\,a\,{e^2}\,\biggl[2a(3- \cos 2\ell) + 4 \cos (\ell+g-{\theta_1})\cr
&\qquad\qquad
-\,3 \cos (3\ell+g-{\theta_1})- \cos (\ell-g+{\theta_1})
\biggr]\,+\,O({e^3})\Biggr\}^{-1/2}\>,\cr}
\eqno  (25)
$$
\noindent
where $\sin v$ and $\cos v$ have been replaced by 
$$\!\!\!\!\!\!\!\!\!\!\!\!\!\!\!\!\!\!\!
\cos v\,=\,-\,e\,+\,2\,\Bigl({{1-{e^2}}\over e}\Bigr)\,
{\sum_{j=1}^{\infty}}\,\cos (j\ell)\>{J_{_j}}(je)\>\>,
\eqno  (26)
$$
\noindent
and
$$
\sin v\,=\,(1-{e^2})^{1/2}\,{\sum_{j=1}^{\infty}}\,
\sin (j\ell)\>\bigl[{J_{_{j-1}}}(je)\,-\,{J_{_{j+1}}}(je)\bigr]\>.
\eqno (27)
$$
\noindent
Here $J_{_j}$ is the Bessel function of order $j$.
From equation (25), $\cal F$ can be written as
$$
{\cal F}\,=\,{{\cal F}^{(0)}}\,+\,e\,{{\cal F}^{(1)}}\,+\,{e^2}\,{{\cal F}^{(2)}}
\,+\,O({e^3})\>\>,
\eqno  (28)
$$
\noindent
where
$$\!\!\!\!\!\!\!\!\!\!\!\!\!\!\!\!\!\!\!\!\!\!\!\!\!\!\!\!\!\!\!\!
\!\!\!\!\!\!\!\!\!\!\!\!\!\!\!\!\!
{{\cal F}^{(0)}}\,=\,a\,{\Bigl[1+{a^2}-2a
\cos (\ell+g-{\theta_1})\Bigr]^{-3/2}}\>
\sin (\ell+g-{\theta_1})\>\>\>,
\eqno  (29)
$$
\vskip  1pt
$$\eqalign {
{{\cal F}^{(1)}}\,=\,&
a\,{\Big[1+{a^2}-2a\cos (\ell+g-{\theta_1})\Bigr]^{-3/2}}
\,\Big[a\sin \ell+\sin (2\ell+g-{\theta_1})\Bigr]\cr
&+\,{3\over 2}\,{a^2}\,
{\Big[1+{a^2}-2a\cos (\ell+g-{\theta_1})\Bigr]^{-5/2}}\cr
&\qquad\qquad \sin (\ell+g-{\theta_1})
\>\Bigl[2a\cos \ell+\cos (2\ell+g-{\theta_1})-
3\cos (g-{\theta_1})\Bigr]\>,\cr}
\eqno  (30)
$$
\vskip 2pt
\noindent
and
\vskip  2pt
$$\eqalign {
{{\cal F}^{(2)}}\,&=\,
{1\over 8}\,a\,
{\Big[1\,+\,{a^2}\,-\,2\,a\,\cos (\ell\,+\,g\,-\,{\theta_1})\Bigr]^{-3/2}}\cr
&\qquad\quad
\Bigl[4a\,\sin 2\ell\,-\,4\sin (\ell+g-{\theta_1})\,+\,
9\sin (3\ell+g-{\theta_1})\,+\,
\sin (\ell-g+{\theta_1})\Bigr]\cr
&-\,{3\over 16}\,{a^2}\,
{\Big[1\,+\,{a^2}\,-\,2\,a\,\cos (\ell\,+\,g\,-\,{\theta_1})\Bigr]^{-5/2}}\cr
&\qquad\qquad
\Bigl[20\,a\,\sin (\ell+g-{\theta_1})-14\,a\, \sin (3\ell+g-{\theta_1})\cr
&\qquad\qquad
+14\,a\,\sin (\ell-g+{\theta_1})+\,16\sin 2(\ell+g-{\theta_1})\cr
&\qquad\qquad
-7\sin 2(2\ell+g-{\theta_1})+
2\,(7\,-4{a^2})\,\sin 2\ell\,-\,\sin 2(g-{\theta_1})\Big]\cr
&+\,{15\over32}\,{a^3}\,
{\Big[1\,+\,{a^2}\,-\,2\,a\,\cos (\ell\,+\,g\,-\,{\theta_1})\Bigr]^{-7/2}}\cr
&\qquad\qquad
\Bigl[2(4{a^2}\,+\,13)\,\sin (\ell+g-{\theta_1})\,+\,
(4{a^2}\,-\,7)\,\sin (3\ell+g-{\theta_1})\cr
&\qquad\qquad
-(4{a^2}\,-\,15)\sin (\ell-g+{\theta_1})\,+\,\sin (5\ell+3g-3{\theta_1})\cr
&\qquad\qquad
+\,9\,\sin (\ell+3g-3{\theta_1})\,+\,
4a\,\sin 2(2\ell+g-{\theta_1})\,-\,16\,a\,\sin 2\ell\cr
&\qquad\qquad
-12\,a\,\sin 2(g-{\theta_1})\,-\,8\,a\,\sin 2(\ell+g-{\theta_1})\,-\,
6\,\sin 3(\ell+g-{\theta_1})\Bigr]\>.\cr}
\eqno (31)
$$
\vskip  7pt

To compute the averaged value of $\bar {\cal F}$, we expand
${[1+{a^2}-2a\cos (\ell+g-{\theta_1})]^{-w}}\,,\,w=3/2,5/2,7/2$,  
using the identity
\vskip  2pt
$$
{(1\,-\,2\,\xi\,\cos \alpha\,+\,{\xi^2})^{-\lambda}}\,=\,
{\sum_{q=0}^{\infty}}\,{C_q^\lambda}\,(\cos \alpha)\,{\xi^q}\>\>\>;\>\>\>
|\xi|<1\>,
\eqno  (32)
$$
\vskip  5pt
\noindent
taking into account that for the outer planet $a>1$. 
In this equation
\vskip 3pt
$$
{C_q^\lambda}(\cos \alpha)\,=\,
{\sum_{h=0}^q}\,{{\Gamma(\lambda+h)\,\Gamma(\lambda+q-h)}\over
{h!\,(q-h)!\,{\Bigl[\Gamma(\lambda)\Bigr]^2}}}\,\cos\bigl[(q-2h)\alpha \bigr]
\eqno  (33)
$$
\vskip 5pt
\noindent
are the Gegenbauer polynomials. To use identity (32) for the calculation of 
$\bar {\cal F}$, one has to set $\alpha = \ell + g - {\theta_1}$ and $\xi = 1/a$.
Simplifying ${{\cal F}^{(0)}},
{{\cal F}^{(1)}}$ and ${\cal F}^{(2)}$ using equation (32), 
one will notice that these quantities will be equal to sum of terms
with a general form of
${e^s}\cos [(q-2h)(\ell+g-{\theta_1})]\sin [\nu\ell+\nu'(g-{\theta_1})]$
where $\nu$ and $\nu'$ are integers.
The harmonic nature of these terms requires that
in an $(n:n')$ resonance, in order for the formula (A7) to
give non-zero values, $(q - 2h)$ has to
be equal to one of the four integers $\pm (\nu \pm n')$ 
and, at the same time, one of the 
four integers $\pm (\nu' \pm n)$ with the same 
arrangement of + and - signs. This immediately implies that 
the general term
${e^s}\cos [(q-2h)(\ell+g-{\theta_1})]\sin [\nu\ell+\nu'(g-{\theta_1})]$
will have non-zero averaged value only if
\vskip  1pt
$$
|\Delta\, \nu| = |\nu - \nu'| = n'-n\,.
\eqno (34)
$$
\vskip  2pt
\noindent
The quantity $|\Delta \nu|$ in the {\it selection rule} (34) is,
in fact, the order of the resonance. An inspection of equation (28)
reveals that in an exterior $(n:n')$ resonance of order $|\Delta \nu|$,
the first fulfillment of the condition (34) by the factor 
$\sin [\nu\ell+\nu'(g-{\theta_1})]$ appears
where $s$ becomes equal to $|\Delta \nu|$. For instance, 
the contribution of expansion (28) to the averaged dynamics of the outer
planet in a resonance of the form $(n:n+1)$ will first appear in its
second term, ${{\cal F}^{(1)}}$,
and the third term of this expansion, ${{\cal F}^{(2)}}$,
will be the first term
with a non-zero averaged value when the system is captured 
in an $(n:n+2)$ commensurability. 

Let us now, just as examples of the first and the second order exterior resonances,
study the averaged system near (1:2) and (1:3) commensurabilities.
In general, after replacing $\ell$ by its equivalent value given by equation (15)
and averaging the results,
the product of $\cos [(q-2h)(\ell+g-{\theta_1})]$ and 
$\sin [\nu\ell\,+\,\nu'(g-{\theta_1})]$
will produce terms that are proportional to
$\sin (n'{\bar \varphi}+ng)$. 
For instance, for a system near a (1:2) resonance, 
\vskip 2pt
$$
{{\bar {\cal F}}_{(1:2)}}\,=\,
{{e_{(1:2)}}\over {a_{(1:2)}^2}}\,{\sigma_{(1:2)}}\,
\sin (2{\bar \varphi} + g)\>,
\eqno  (35)
$$
\vskip 2pt
\noindent
and for a (1:3) resonance,
\vskip  2pt
$$
{{\bar{\cal F}}_{(1:3)}}\,=\,{{e_{(1:3)}^2}\over {4{a_{(1:3)}^2}}}\,
\Bigl[{\sigma_{(1:3)}^{(3/2)}}\,+\,3{\sigma_{(1:3)}^{(5/2)}}\,+\,
{15\over {a_{(1:3)}^2}}{\sigma_{(3:1)}^{(7/2)}}\Bigr]\,
\sin (3{\bar \varphi} + g)\>.
\eqno  (36)
$$
\vskip 3pt
\noindent
In these equations,
\vskip 2pt
$$\!\!\!\!\!\!\!\!\!\!\!\!\!\!\!\!\!\!\!\!\!\!\!\!\!\!
\!\!\!
{\sigma_{(1:2)}}\,=\,{\sum_{h=0}^\infty}\>{\Biggl[{{\Gamma ({3\over 2}+h)}\over
{{a_{(1:2)}^h}\>h!\>\Gamma({3\over 2})}}\Biggr]^2} 
\Biggl\{\,1\,+\biggl({{2h+3}\over {h+1}}\biggr)\>
\biggl[\,1\,-\,{3\over {4{a_{(1:2)}^2}}}\>\biggl({{2h+5}
\over {h+2}}\biggr)\biggr]\Biggr\}\>,
\eqno  (37)
$$
$$\!\!\!\!\!\!\!\!\!\!\!\!\!\!\!\!\!\!\!\!\!\!\!\!\!\!\!\!\!\!
{\sigma_{(1:3)}^{(3/2)}}\,=\,{\sum_{h=0}^\infty}\>
{\Biggl[{{\Gamma ({3\over 2}+h)}\over
{{a_{(1:3)}^h}\>h!\>\Gamma({3\over 2})}}\Biggr]^2} \,
\Biggl\{{9\over 2}\,+\,\biggl({{2h+3}\over {h+1}}\biggr)\,
\biggl[1\,+\,{1\over {8{a_{(1:3)}^2}}}
\Bigl({{2h+5}\over {h+2}}\Bigr)\biggr]\Biggr\}\>,
\eqno  (38)
$$
$$
\!\!\!\!\!\!\!\!\!\!\!\!\!\!\!\!\!\!\!\!\!\!\!\!\!\!\!\!\!\!
\!\!\!\!\!\!\!\!\!\!\!\!\!\!\!\!\!\!\!\!\!
\eqalign{
{\sigma_{(1:3)}^{(5/2)}}\,=\,{\sum_{h=0}^\infty}\>
&{\Biggl[{{\Gamma ({5\over 2}+h)}\over
{{a_{(1:3)}^h}\>h!\>\Gamma({5\over 2})}}\Biggr]^2} \,
\Biggl\{{7\over 2}\,+\Bigl[1-{7\over {8{a_{(1:3)}^2}}}\Bigr]\,
\Bigl({{2h+5}\over {h+1}}\Bigr)\cr
&-{1\over {8{a_{(1:3)}^2}}}\,
\biggl[7\,+\,{1\over {4{a_{(1:3)}^2}}}\Bigl({{2h+9}\over {h+3}}\Bigr)
\biggr]
{{(2h+7)(2h+5)}\over {(h+2)(h+1)}}\Biggr\}\>,\cr}
\eqno  (39)
$$
$$\!\>\>\>
\eqalign{
\,{\sigma_{(1:3)}^{(7/2)}}={\sum_{h=0}^\infty}\>&{\Biggl[{{\Gamma ({7\over 2}+h)}\over
{{a_{(1:3)}^h}\>h!\>\Gamma({7\over 2})}}\Biggr]^2} 
\Biggl\{{1\over 2}\Bigl[{a_{(1:3)}^2}-{7\over 4}\Bigr]-
{3\over 4}\Bigl({{2h+7}\over {h+1}}\Bigr) \cr
&+\Bigl[{{1+15{a_{(1:3)}^2}-4{a_{(1:3)}^4}}\over {32{a_{(1:3)}^2}}}\Bigr]
{{(2h+9)(2h+7)}\over {(h+2)(h+1)}}\cr
&+{3\over {16{a_{(1:3)}^2}}}
\biggl[1-{3\over {8{a_{(1:3)}^2}}}\Bigl({{2h+13}\over {h+4}}\Bigr)
\biggr]
{{(2h+11)(2h+9)(2h+7)}\over {(h+3)(h+2)(h+1)}}\Biggr\}.\cr}
\eqno  (40)
$$
\vskip  4pt
Because $\bar {\cal F}$ is proportional to $\sin (n'{\bar \varphi}+ng)$, 
equation (23) can be viewed as the equation 
of a mathematical pendulum with a potential function proportional to
$\cos (n'{\bar \varphi}+ng)$ (see equation (A12)). 
Such a pendulum, with its harmonic potential, is a characteristic 
of the first-order partially averaged system near a resonance 
where $e_{(n:n')}$ is considered to be constant
(Sanders $\&$ Verhulst 1985; Lichtenberg $\&$ Lieberman 1992;
Wiggins 1996; Haghighipour 1999).
For instance, for the cases of (1:2) and (1:3) resonances,
\vskip 1pt
\vbox{
\vskip  30pt
\noindent
\hskip  27pt
\epsfbox {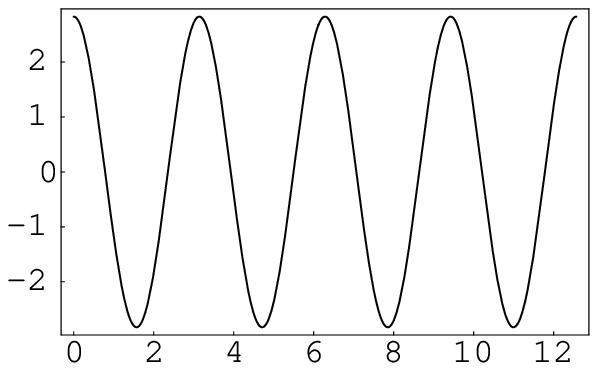}
\hskip 10pt
\epsfbox {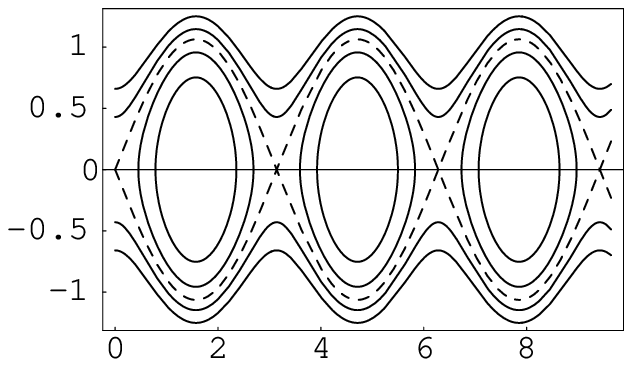}
\hfill
\vskip 20pt
\noindent
\hskip  17pt
\epsfbox{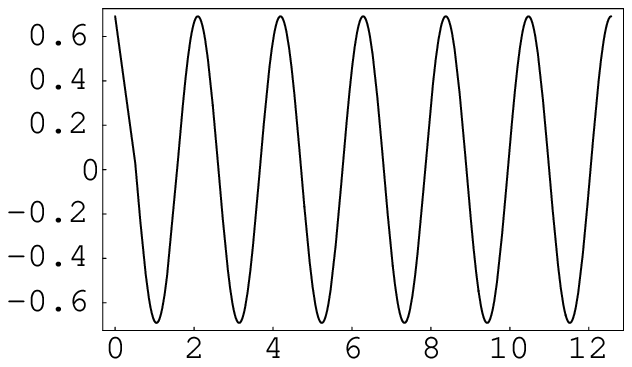}
\hskip  8pt
\epsfbox{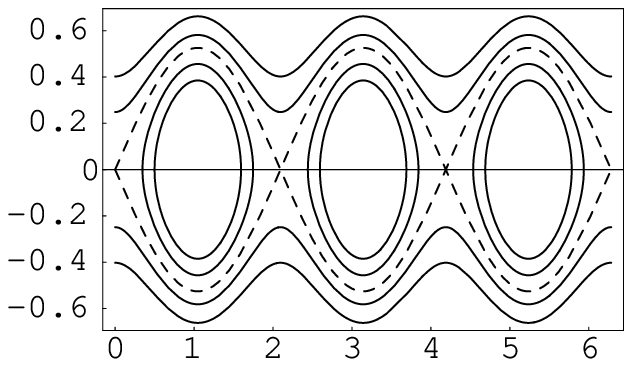}
\hfill}
\vskip  -3.045in
\vbox{
\noindent
${{\cal U}_{(1:2)}}({\bar \varphi})$
\hskip 2.5in
${\dot {\bar \varphi}}$
\hskip 0.459in
$\bullet$
\hskip 0.535in
$\bullet$
\hskip 0.535in
$\bullet$}
\vskip 1pt
\vskip -11pt
\hskip 3.37in
${\cal C}\qquad\!{\cal S}$
\vskip 1.6in
\vbox{
\noindent
${{\cal U}_{(1:3)}}({\bar \varphi})$
\hskip 2.5in
${\dot {\bar \varphi}}$
\hskip 0.435in
$\bullet$
\hskip 0.55in
$\bullet$
\hskip 0.55in
$\bullet$}
\vskip 1pt
\vskip -10.1pt
\hskip 3.32in
${\cal C}\qquad{\cal S}$
\vskip 40pt
\noindent
\hskip 1.5in
${\bar \varphi}$(rad)
$\qquad\qquad\qquad\qquad\qquad\qquad\qquad\qquad 
{\bar \varphi}$(rad)
\vskip  0.01in
\vbox{
\baselineskip=\normalbaselineskip
\vskip 10pt
\noindent
{\smallrmb Figure 1.} {\smallrm Graphs of the potential function 
${{\cal U}_{(1:2)}}({\bar \varphi})$ (top left) and its phase diagram (top right)
and the potential function ${{\cal U}_{(1:3)}}({\bar \varphi})$ (bottom left)
with its associated phase diagram (bottom right)
for the system studied by Haghighipour (1999) at (1:2) and (1:3) resonances. 
The scale on all vertical axes is 0.01 . The origins on the horizontal axes of the
graphs of the (1:2) resonance have been shifted by -0.44(rad) and the corresponding origins
of the graphs of the (1:3) resonance have been shifted by -1.7(rad). }}
\vskip .2in
$$
{{\cal U}_{(1:2)}}({\bar \varphi})\,=\,{{3{e_{(1:2)}}}\over 
{2{a_{(1:2)}^4}}}\,
{\sigma_{(1:2)}}\,\cos (2{\bar \varphi} + g)\,+\, Constant,
\eqno (41)
$$
\noindent
and 
$$
{{\cal U}_{(1:3)}}({\bar \varphi})\,=\,{{e_{(1:3)}^2}\over 
{4{a_{(1:3)}^4}}}
\Bigl[{\sigma_{(1:3)}^{(3/2)}}\,+\,3\,{\sigma_{(1:3)}^{(5/2)}}\,+\,
{{15}\over {a_{(1:3)}^2}}{\sigma_{(1:3)}^{(7/2)}}\Bigr]\,
\cos (3{\bar \varphi} + g)\>+\>
Constant.
\eqno  (42)
$$
\vskip  3pt
\noindent
Figure 1 shows the graphs of these two potential functions with their
corresponding phase diagrams. In producing these graphs, 
the numerical values of the orbital eccentricity 
\noindent
{\centerline{
\hskip  -5pt
\epsfbox {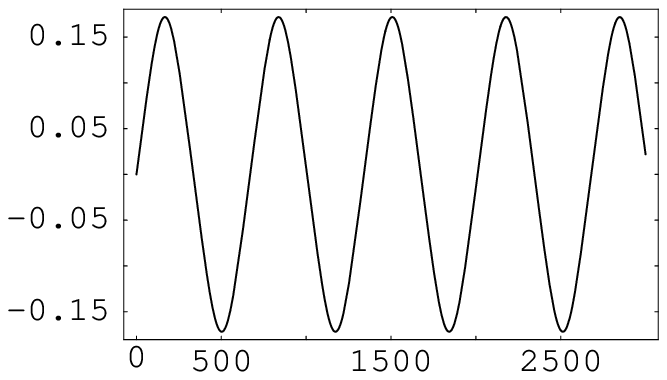}
\hskip 20pt
\epsfbox {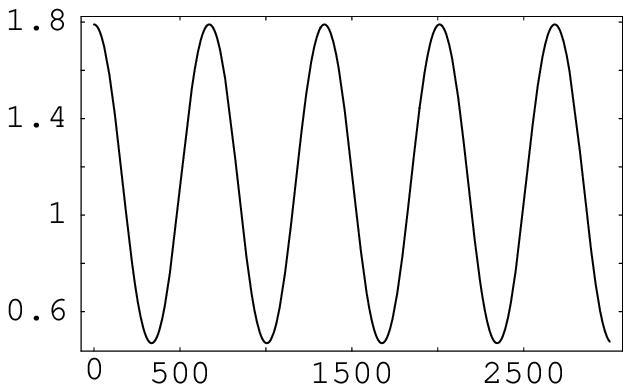}}}
\vskip 20pt
{\centerline{
\epsfbox{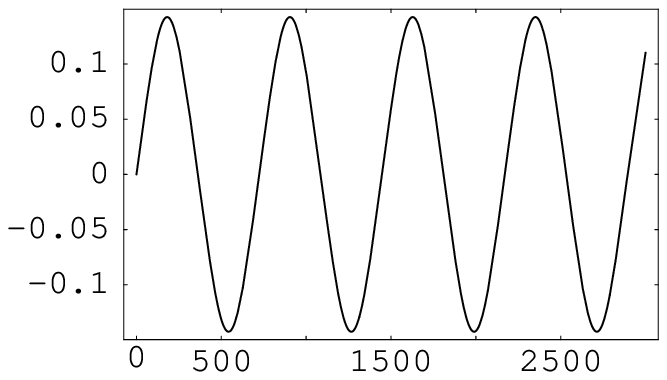}
\hskip  15pt
\epsfbox{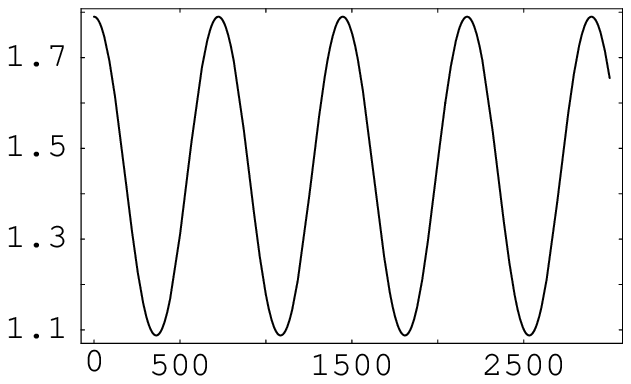}}}
\vskip  -2.92in
\vbox{
\noindent
${{\bar D}_{(1:2)}}
\qquad\qquad\qquad \qquad\qquad\qquad\qquad\qquad\qquad
{{\bar \varphi}_{(1:2)}}$}
\vskip 1.62in
\vbox{
\noindent
${{\bar D}_{(1:3)}}
\qquad\qquad\qquad \qquad\qquad\qquad\qquad\qquad\qquad
{{\bar \varphi}_{(1:3)}}$}
\vskip 45pt
\noindent
\hskip 1.3in
Time $({\hat t})$ 
$\qquad\>\>\qquad\qquad\qquad\qquad\qquad\qquad\qquad$ 
Time $({\hat t})$
\vskip  0.1in
\vbox{
\baselineskip=\normalbaselineskip
\vskip 10pt
\noindent
{\smallrmb Figure 2.} {\smallrm Graphs of ${\bar D}({\hat t})$ and ${\bar
\varphi}({\hat t})$  for the systems of figure 1, partially averaged  
to the first order.}}
\vskip .3in
\noindent
and the semimajor axis of $P_2$ have been taken from
the restricted three-body system of Haghighipour (1999) at (1:2)
and (1:3) resonances.

The harmonic nature of the potential function of the pendulum (23)
indicates that this pendulum can be in three dynamical states;
stable equilibrium corresponding to the minimum of the potential
function (centers $\cal C$ on the phase diagrams),
unstable equilibrium corresponding to the maximum of the potential
function (saddle points $\cal S$) or, an oscillatory 
(librational) motion around the stable equilibrium 
(the orbits inside the {\it separatrix}, the dashed orbit
that passes through the saddle point $\cal S$).
The resonance lock phenomenon is geometrically depicted by
these librational motions.

The oscillatory variations in values
of $\bar \varphi$ create a harmonic behavior for the action
variable $\bar D$. From equations (22), for a system at an $(n:n')$
resonance where $\bar {\cal F}$ is proportional to 
$\sin (n'{\bar \varphi}+ng)\>,\, {\bar D}$ will be proportional to
$\cos (n'{\bar \varphi}+ng)$. Figure 2 shows the graphs of $\bar D$ and
$\bar \varphi$ against time for the systems of figure 1. As mentioned in section 3,
$D$ is the measure of changes in the action variable $L$ or, in other
words, an indication of the variations of the semimajor axis of the outer
planet from its resonant value. From equation (A14),
the width of the resonance band within which
the semimajor axis of the outer planet varies around its resonant value
is limited by the height of the separatrix and to the first order
of perturbation can be written as
\vskip  2pt
$$
\Delta {a_{(n:n')}}\,=\,4\,{\Bigl[{2\over 3}\,\mu\,
{a_{(n:n')}^3}\,\Delta\,{{\cal U}_{(n:n')}}({\bar \varphi})\Bigr]^{1/2}}
\eqno (43)
$$
\vskip  5pt
\noindent
where $\Delta\,{{\cal U}_{(n:n')}}({\bar \varphi})$ is equal to the
difference between the maximum and the minimum values of 
${{\cal U}_{(n:n')}}({\bar \varphi})$. 

It is necessary to mention that the procedure
presented here for expansion of $[1+{a^2}-2a\cos (\ell+g-{\theta_1})]^{-w}$
using Gegenbauer polynomials, is not valid for a (1:1) resonance.
At this state,  from equation (13), the
Keplerian period of the outer planet becomes nearly equal to $T_1$. 
That means, ${a_{(1:1)}}\simeq 1$ and expansion (32) is no longer 
applicable. However, it is still possible
to use the method of averaging to study the dynamics of
the outer planet near a (1:1) commensurability.
The {\it external Hamiltonian} $H$, in this case, must 
be studied in its entirety as given by equation (25). 
Expanding $H$ to the second
order in eccentricity and integrating $\cal F$ in the vicinity
of the (1:1) resonance, we have
\vskip  5pt
$$\eqalign{
{{\bar {\cal F}}_{(1:1)}}\,=\,&
{1\over 2}\,{a_{(1:1)}}\,[2\,-\,{e_{(1:1)}^2}]\,\sin ({\bar \varphi}\,+\,g)\,
{\biggl[1+{a_{(1:1)}^2}-2{a_{(1:1)}} \cos ({\bar \varphi} + g)\biggr]^{-3/2}}\cr
&\quad
-\,3\,{a_{(1:1)}^2}\,{e_{(1:1)}^2}\,\sin 2 ({\bar \varphi}\,+\,g)\,
{\biggl[1+{a_{(1:1)}^2}-2{a_{(1:1)}} \cos ({\bar \varphi} + g)\biggr]^{-5/2}}\cr
&\quad
+\,{{45}\over 4}\,{a_{(1:1)}^3}\,{e_{(1:1)}^2}\,{\sin ^3} ({\bar \varphi}\,+\,g)
{\biggl[1+{a_{(1:1)}^2}-2{a_{(1:1)}} \cos ({\bar \varphi} + g)\biggr]^{-7/2}}\,.\cr}
\eqno  (44)
$$
\vfill
\eject
\hskip  7pt
\epsfbox {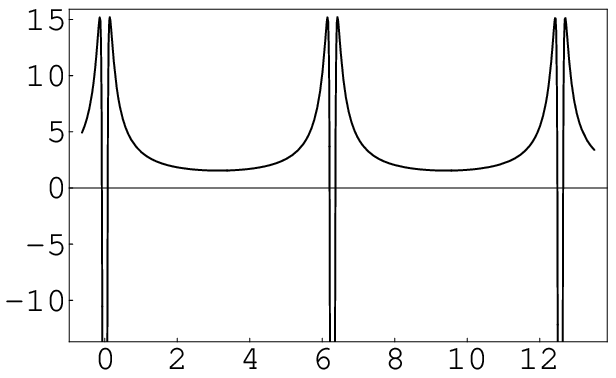}
\hskip 5pt
\epsfbox {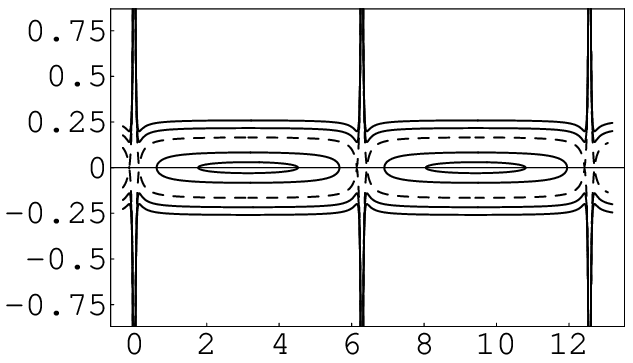}
\vskip -1.06in
\noindent
${{\cal U}_{(1:1)}}({\bar \varphi})
\qquad\qquad\qquad \qquad\qquad\qquad\qquad\qquad\qquad\>
{\dot {\bar \varphi}}$
\vskip .62in
\hskip 1.35in
$\bar \varphi$(rad)
\hskip 2in
$\bar\varphi$(rad)
\vskip 5pt
\vbox{
\baselineskip=\normalbaselineskip
\noindent
{\smallrmb Figure 3} {\smallrm Graphs of the potential 
function ${{\cal U}_{(1:1)}}({\bar \varphi})$ (left) and its phase diagram 
(right) against $\bar \varphi$.}}
\vskip 20pt
\noindent
The potential function associated 
with the first-order partially averaged system 
is given by
$$\eqalign{
{{\cal U}_{(1:1)}}({\bar \varphi)}\,=\,&
{3\over {a_{(1:1)}^2}}\,
{\biggl[1+{a_{(1:1)}^2}-2{a_{(1:1)}} \cos ({\bar \varphi} + g)\biggr]^{-1/2}}\cr
&\quad
-\,{{21}\over {2{a_{(1:1)}}}}\,{e_{(1:1)}^2}\,\cos  ({\bar \varphi}\,+\,g)\,
{\biggl[1+{a_{(1:1)}^2}-2{a_{(1:1)}} \cos ({\bar \varphi} + g)\biggr]^{-3/2}}\cr
&\quad
+\,{{27}\over 4}\,{e_{(1:1)}^2}\,{\sin ^2} ({\bar \varphi}\,+\,g)
{\biggl[1+{a_{(1:1)}^2}-2{a_{(1:1)}} \cos ({\bar \varphi} + g)\biggr]^{-5/2}}\,.\cr}
\eqno  (45)
$$
\vskip 20pt
\noindent
Figure 3 shows the graph of this potential 
for $g=0^\circ$. The fact that 
$d{{\cal U}_{(1:1)}}({\bar \varphi})/d{\bar \varphi}$
shows slight deviations from zero near the points of stable 
equilibrium is an indication of a slow-frequency librational motion
in that neighborhood. To see this, let us apply
this analysis to the system of Sun-Jupiter-Trojan asteroid.
Located at the $L_4$ and $L_5$ Lagrangian points of Jupiter's orbit,
Trojan asteroids are in a near $(1:1)$ resonance with
Jupiter and have a librational motion with a period of approximately 
148 years (Brown $\&$ Shook 1964; Fleming $\&$ Hamilton 2000).
Substituting for $a$ and $e$ in equation (45) by the values 
of the semimajor axis and the eccentricity of Trojan asteroids and
Taylor expanding ${\cal U}_{(1:1)}({\bar \varphi})$ 
around its stable equilibrium,
one will obtain a librational period of approximately 110 years.
The difference between this period and the 148 years reported in 
the references above can be attributed to several factors such as 
neglecting the gravitational effect of Saturn, restricting Jupiter
to stay on a circular orbit and also to the coupling between 
$\varphi$ and the argument of the pericenter $g$. Studies are currently
underway to extend the partially averaged equations of this system to the
second order of perturbation where $\varphi$ and $g$ decouple.
A closer value for the librational period above is expected in this case.

The first-order partially averaged system near a resonance,
presented by equation (23) is, in general, Hamiltonian. 
It portrays the phenomenon of capture in a resonance as 
librational motion of a pendulum. However,
with respect to an arbitrary perturbation, 
this Hamiltonian system is structurally unstable. 
In an actual system, in order to be able to 
draw conclusions on long-term behavior of quantities 
such as orbital eccentricity of the outer planet, its 
angular momentum and also the precession of its orbit,
it is necessary to extend this analysis to higher
orders of the perturbation parameter $\mu^{1/2}$. 
Such an extension will allow for the time variation of the angle $g$ to be
taken into consideration. This will render the system of equations 
(18) to (21) in
a set of equations with two angular variables, $\varphi$ and $g$. 
In order to be able to apply
the method of partial averaging near a resonance to this system, it is
then necessary to
introduce an averaging transformation that renders the equations of the
system in a form that to the first order of perturbation, it 
becomes automatically equivalent to the
first-order partially averaged system at resonance. The second-order
partially averaged system is then obtained by averaging those equations using
formula (A7). Such studies are currently in preparation for publication.
\vskip 10pt
\noindent
{\bigrmsixteen {5. $\>$ SUMMARY AND CONCLUDING REMARKS}}
\vskip  10pt 

Application of the method of partial averaging to the study of
the dynamics of the outer body of a restricted three-body system 
while captured in a resonance has been presented here. 
Analysis of the first-order partially averaged system near a
resonance has revealed that the equations of motion of the
outer planet, averaged over fast periodic motion at resonance,  
resembles a mathematical pendulum. Such a pendulum
analogy can also be found in the comprehensive study of orbital
resonances among planetary satellites by Peale (1986)
and also in the comprehensive study of the dynamical behavior of a test 
particle near an interior as well as an exterior resonance in a restricted
three-body system by Winter and Murray (1997a$\&$b).
 
In the analysis presented here, the driving force of the pendulum-like
first-order partially averaged system is obtained from the 
{\it external Hamiltonian}
$H$ (equation (24)) which involves the gravitational effect of the inner
planet on the dynamics of the outer one. In writing the dynamical
equations of the outer planet in terms of the Delaunay variables,
this force appears as $F_r$ and $F_\theta$ 
in equations (8) to (11). It is important to mention that the form
of these equations (i.e., equations (8) to (11))
is quite general and independent of the
physical nature of the perturbation.
In a system where the perturbations are non-Hamiltonian,
extra terms will be added to the functions
$F_r$ and $F_\theta$ as well as equations (16) and (17). 
However, equations (8) to (11) will keep their general form
(Haghighipour 1999, 2000).
The procedure presented here regarding the application of the
partial averaging technique and the analysis of the pendulum-like
equation are also quite general and can be applied 
to non-Hamiltonian systems in a similar fashion.
In fact, one of the most important features 
of the method of partial averaging near a resonance is that it 
presents a general analytical procedure that
is equally applicable to both
Hamiltonian and non-Hamiltonian systems. As examples
of the systems where the method of partial averaging 
near a resonance has been used
in conjunction with non-Hamiltonian perturbations, 
we refer the reader to  Chicone, Mashhoon
$\&$ Retzloff (1997a$\&$b) and Haghighipour (1999).

As mentioned in section 4,  the first-order partially averaged system
near a resonance presents the first step in utilizing the method of
averaging
in the analytical study of the dynamics of a resonance-locked 
system. To this order of approximation, the argument of the pericenter
$g$ was assumed to be constant.
In order to obtain a more comprehensive picture of the dynamics 
of a system near a resonance and the roles that perturbative effects
play in its stability, 
one has to extend this analysis to higher orders of the perturbation
parameter $\mu^{1/2}$. Such an extension is necessary to assure decoupling of
the angles $\varphi$ and $g$ (e.g., see Haghighipour 2000).
At that stage, one can apply the
averaging technique presented here to the dynamics of the 
system at the second order of perturbation
by averaging those equations over the fast-changing
angular variable $\varphi$ and studying the pendulum-like equation of the
angular variable $g$. Such studies are currently underway for  the case of (1:1)
resonance and their applications to Trojan asteroids.

The choice of a restricted system as presented in this study 
was merely to focus attention on the method of partial averaging and
its capabilities as another approach to analytical analysis of the dynamics 
of a system near a resonance. The analysis presented here
is quite general and can be applied to any dynamical system at resonance.
One can apply such analysis
to the systems at interior resonances by setting
$\xi=a$ in equation (32) and changing the {\it selection rule} (34) to
$$
|\Delta \nu|\,=\,n\,-\,n'\>.
\eqno  (46)
$$
\noindent
An implication of this {\it selection rule} can be found in a recent
paper by Michtchenko $\&$ Ferraz-Mello (2001) on analytical modeling
of the Jupiter-Saturn system near their (5:2) resonance. 
They show that at the lowest order, 
the contribution of the resonant part of the disturbing function
appears as the third power of eccentricity, 
a result that is also implied by the {\it selection rule} (46).

Other interesting cases for application of the partial averaging technique are
the study of the stability of the extrasolar planetary system 
Gliese 876 (Marcy et al. 2001) where its two planets are 
locked in a near (2:1)
commensurability, the study of dynamical stability of P-type binary planetary
systems where recent numerical integrations by Holman $\&$
Wiegert (1999) have indicated the existence of islands of instabilities
for eccentric binaries at $(n:1),n>3$, resonances and also the study of
the dynamical evolution of pulsar planetary systems such as PSR B1257 +12
(Konacki, Maciejewski $\&$ Wolszczan 1999) and PSR B1620-26
(Thorsett et al. 1999; Ford et al. 2000; Ford, Kozinsky $\&$ Rasio 2000;
Rasio 2001). It appears that gravitational radiation reaction
may play a vital role in the dynamical evolution of these system 
(Chicone, Mashhoon $\&$ Retzloff 1996 a$\&$b, 1997 a$\&$b, 1999, 2000) 
\vskip  15pt
\noindent
{\bigrmsixteen {ACKNOWLEDGEMENT}}
\vskip  5pt

I would like to thank B. Mashhoon, C. Chicone, F. Varadi, D. Hamilton
and M. Murison for their fruitful comments and
stimulating discussions. I am especially thankful to B. Mashhoon
and C. Chicone for critically reading the original manuscript.
I would also like to thank
the Department of Physics and Astronomy at the University of 
Missouri-Columbia for their warm hospitality during the course
of this study.
\vskip  20pt
\noindent
{\bigrmsixteen {REFERENCES}}
\vskip  5pt
\noindent
Andoyer M. H., 1903, Bull. Astron., 20, 321
\vskip  1pt
\noindent
Arnold V. I., Kozlov V. V., Neishtadt A.I., 1988, 
Dynamical Systems III. 
\vskip 1pt
Springer-Verlag, New York
\vskip  1pt
\noindent
Brown E. W., Shook C. A., 1964, 
Planetary Theory. Dover, New York, p. 255
\vskip  1pt
\noindent
Cucu-Dumitrescu C., Selaru D., 1998, Celest.Mech.Dynamic.Astron., 69, 255
\vskip  1pt
\noindent
Chicone C., Mashhoon B., Retzloff D. G., 1996a,
Ann. Inst. Henri Poincar\'e, Phys. 
\vskip  1pt
Th\'eorique, 64, 87
\vskip 1pt
\noindent
Chicone C., Mashhoon B., Retzloff D. G., 1996b,
J. Math. Phys., 37, 3997
\vskip 1pt
\noindent
Chicone C., Mashhoon B., Retzloff D. G., 1997a,
Classical Quantum Gravity, 14, 699
\vskip  1pt
\noindent
Chicone C., Mashhoon B., Retzloff D. G., 1997b,
Classical Quantum Gravity, 14, 1831
\vskip  1pt
\noindent
Chicone C., Mashhoon B., Retzloff D. G., 1999, 
Classical Quantum Gravity, 16, 507
\vskip  1pt
\noindent
Chicone C., Mashhoon B., Retzloff D. G., 2000,
J. Phys. A: Math. Gen., 33, 513
\vskip 1pt
\noindent
Dermott S.F., Murray C.D., 1983, Nature, 319, 201
\vskip  1pt
\noindent
Fleming H. J., Hamilton D. P., 2000, Icarus, 148, 479
\vskip  1pt
\noindent
Ferraz-Mello S., 1997, Celest.Mech.Dynamic.Astron., 66, 39
\vskip  1pt
\noindent
Ford E. B., Joshi K. J., Rasio F. A., Zbarsky B., 2000, ApJ, 528, 336
\vskip  1pt
\noindent
Ford E. B., Kozinsky B., Rasio F. A., 2000, ApJ, 535, 385
\vskip  1pt
\noindent
Greenspan B. D., Holmes P. J., 1983, in 
Barenblatt G, Iooss G and Joseph D. D., eds, 
\vskip  1pt
Nonlinear Dynamics and Turbulence, Pitman, London, p.172
\vskip  1pt
\noindent
Grebenikov E. A., Ryabov Yu. A., 1983, Constructive Methods in the
Analysis of
\vskip 1pt
Nonlinear Systems. Mir Publishers, Moscow
\vskip  1pt
\noindent
Guckenheimer J., Holmes P. J., 1983, Nonlinear Oscillations,
Dynamical Systems, 
\vskip  1pt
and Bifurcation of Vector Fields. Springer-Verlag,
New York
\vskip 1pt
\noindent
Haghighipour N., 1999, MNRAS, 304, 185
\vskip 1pt
\noindent
Haghighipour N., 2000, MNRAS 316, 845
\vskip  1pt
\noindent
Henrard J., Lema\^\i tre A., 1983, Celest.Mech., 30, 197
\vskip  1pt
\noindent
Holman M. J., Wiegert P. A., 1999, AJ, 117, 621
\vskip  1pt
\noindent
Konacki M., Maciejewski A. J., Wolszczan. A., 1999, ApJ, 513, 471
\vskip  1pt
\noindent
Laughlin G., Chambers J. E., 2001, ApJL, 551, L109
\vskip  1pt
\noindent
Lee M. H., Peale S. J., 2001, Talk Presented at the 32nd Annual Meeting
\vskip 1pt
of the Division on Dynamical Astronomy, Houston TX
\vskip  1pt
\noindent
Lichtenberg A. J., Lieberman M. A., 1992, Regular and Chaotic Dynamics.
\vskip  1pt
Springer-Verlag, New York, p.29
\vskip1pt
\noindent
Lissauer J. J., Rivera E. J., 2001, ApJ
(to appear in Vol 554 on June 10)
\vskip 1pt
\noindent
Marcy G. W., Butler R. P., Fischer D., Vogt S. S.,
Lissauer J. J., Rivera E. J., 
\vskip  1pt
2001, ApJ (to appear on Volume 556) 
\vskip  1pt
\noindent
Melnikov V. K., 1963, Trans. Moscow Math. Soc., 12, 1
\vskip 1pt
\noindent
Michtchenko T.A., Ferraz-Mello S., 2001, Icarus, 149, 357
\vskip  1pt
\noindent
Murray N., Paskowitz M., Holman M. J., 2001, astro-ph/0104475
\vskip  1pt
\noindent
Peale S. J., 1986, in  Burns J. A. and 
Matthews M. S., eds, Satellites. Univ. Arizona 
\vskip  1pt
Press, Tucson, p.159
\vskip  1pt
\noindent
Poincar\'e H., 1902, Bull. Astron., 19, 289
\vskip  1pt
\noindent
Rasio, F. A., 2001, in Podsiadlowski P. et al., eds,
ASP Conference Series, 
\vskip  1pt
Evolution of Binary and Multiple Star Systems,
\vskip  1pt
\noindent
Sanders J. A., Verhulst F., 1985, Averaging Methods
in Nonlinear Dynamical Systems. 
\vskip  1pt
Springer-Verlag, New York
\vskip  1pt
\noindent
Snellgrove M. D., Papaloizou J. C. B., Nelson R. P., 2001, astro-ph/0104432
\vskip  1pt
\noindent
Thorsett S. E., Arzoumanian Z., Camilo F., Lyne A. G.,
1999, ApJ, 523, 763
\vskip  1pt
\noindent
Wiggins S., 1996, Introduction to Applied Nonlinear Dynamical Systems
and Chaos.
\vskip  1pt
Springer-Verlag, New York, p.143
\vskip  1pt
\noindent
Winter O. C., Murray C. D., 1997a, A$\&$A, 319, 290
\vskip  1pt
\noindent
Winter O. C., Murray C. D., 1997b, A$\&$A, 328, 399
\vskip  35pt
\noindent
{\bigrmsixteen APPENDIX  A:  
METHOD OF PARTIAL AVERAGING NEAR A RESONANCE}
\vskip  15pt

A short introduction to the method of partial averaging near a resonance
as used in this article is 
presented here. For more details on this technique, the reader is referred
to Sanders $\&$ Verhulst (1985), Wiggins (1996),
Haghighipour (1999, 2000) and the references therein.

Partial averaging near a resonance is based on the
application of the Method of Averaging
to the dynamical equations of a system 
in the vicinity of its resonant state
(Sanders $\&$ Verhulst 1985; Wiggins 1996). 
These equations are usually written
in one of the several standard forms (Sanders $\&$ Verhulst
1985). In celestial mechanics, it is customary to write 
the dynamical equations of the
system in terms of action-angle variables. 

Consider a perturbation system with an action variable 
$\cal B$ and an angular variable $\beta$ such that
\vskip  1pt
$$
{\dot {\cal B}}\,=\,\varepsilon\,{\cal M} ({\cal B}\,,\,\beta\,,\,t\,,\,\varepsilon)\>,
\eqno  (A1)
$$
\noindent
and
$$
{\dot \beta}\,=\,{\omega_0} ({\cal B})\,+\,\varepsilon\,{\cal Q} 
({\cal B}\,,\,\beta\,,\,t\,,\,\varepsilon)\>,
\eqno  (A2)
$$
\vskip 2pt
\noindent
where $\cal M$ and $\cal Q$ are periodic in time with 
period ${\cal T}$ and $\omega_0$ is the frequency of the unperturbed
$(\varepsilon=0)$ system. 
At resonance $\cal T$ and $\omega_0$ are related as
$$
l\,{\omega_0}\,=\,l'\,{\omega_{\cal T}}\>,
\eqno  (A3)
$$
\vskip  2pt
\noindent
where $\omega_{\cal T}$ is the angular frequency associated 
with $\cal T$ and $l$ and $l'$
are positive integers. One can show that in the
vicinity of the resonance state (A3), equations
(A1) and (A2) can be written as
(Wiggins 1996; Haghighipour 2000)
\vskip 2pt
$$\!\!\!\!\!\!\!\!\!\!\!\!\!\!\!\!\!\!
\!\!\!\!\!\!\!\!\!\!\!\!\!\!\!\!\!\!\!
{\dot{\cal E}}\,=\,{\varepsilon^{1/2}}\,{\cal M} 
({{\cal B}_0},{\beta},t)\,,+\,
\varepsilon\, {\cal E}\,{{\partial {\cal M}}\over 
{\partial {\cal B}}}({{\cal B}_0}, \beta,t)\,+\,
O({\varepsilon^{3/2}})
\eqno (A4)
$$
\noindent
and
\vskip  2pt
$$
{\dot \Theta}\,=\,{\varepsilon^{1/2}}\,{\cal E}\,
{{\partial {\omega_0}}\over {\partial{\cal B}}}\,({{\cal B}_0})\,+\,
\varepsilon\,\Bigl[{\cal Q}({{\cal B}_0}, \beta,t)\,+\,
{1\over 2}\,{{\cal E}^2}\,
{{{\partial^2} {\omega_0}}\over {\partial {{\cal B}^2}}}({{\cal B}_0})\Bigr]
\,+\,O({\varepsilon^{3/2}})\,,
\eqno (A5)
$$
\vskip 2pt
\noindent
where
$$
{\cal E}={\varepsilon^{-1/2}}({\cal B}-{{\cal B}_0})
\qquad,\qquad
\Theta=\beta-{\omega_0}({{\cal B}_0})t\,, 
\eqno  (A6)
$$
\vskip  2pt
\noindent
represent deviations of $\cal B$
and $\beta$ from their resonant values ${\cal B}_0$ and 
${\omega_0}({{\cal B}_0})t$, respectively. 
Equations (A6) are, indeed, the necessary
transformations for writing equations (A1) 
and (A2) in the 
neighborhood of the $(l:l')$ resonance.
The averaged dynamics of the system at this neighborhood is
obtained by averaging equations (A4) and (A5) using the averaging
integral
$$
{\bar {\cal R}}\,=\,{1\over {l{\cal T}}}\,{\int_0^{l{\cal T}}}\,
{\cal R}\bigl[{{\cal B}_0}\,,\,{\omega_0}({{\cal B}_0})t+
{\Theta}\,,\,t\bigr]\,dt\>.
\eqno  (A7)
$$

The first order partially averaged system near $(l:l')$ resonance
is obtained by neglecting the $O(\varepsilon)$ terms
in equations (A4) and (A5) and is given by
\vskip 2pt
$$
{\dot{\bar{\cal E}}}\,=\,{\varepsilon^{1/2}}\,{\bar {\cal M}} 
({{\cal B}_0}\,,\,{\bar \Theta})\>,
\eqno (A8)
$$
\noindent
and
$$
{\dot{\bar \Theta}}\,=\,{\varepsilon^{1/2}}\,{\bar {\cal E}}\,
{{\partial {\omega_0}}\over {\partial{\cal B}}}\,({{\cal B}_0})\>,
\eqno (A9)
$$
\vskip  10pt
\noindent
where the overbar indicates an averaged quantity.
According to the Principle of Averaging (Saners $\&$ Verhulst 1985;
Wiggins 1996; Chicone, Mashhoon $\&$ Retzloff 1997a; Haghighipour 1999)
the dynamics of the system of equations (A1) and (A2) can be approximated
by the dynamics of the system (A8) and (A9) during the time interval
${\varepsilon^{-1/2}}t$. It is necessary to emphasize that in order 
to be able to make such an approximation, it is required by the Principle
of Averaging, that the main dynamical system (i.e., equations (A1) and (A2))
 to have only one angular variable.
Extension of the method of averaging to the systems with two
or more angular variables can be found in the works of
Grebenikov $\&$ Ryabov (1983), Arnold, Kozlov $\&$ Neishtadt (1988) and also in a recent
paper by Cucu-Dumitrescu $\&$ Selaru (1998) on study of 
the equations of motion around an oblate planet. 
In the first order of perturbation, however, 
such an extension is not necessary.

Introducing $\cal H$, as
\vskip 2pt
$$
{\cal H}\,=\,{\varepsilon^{1/2}}\,
\biggl[{1\over 2}\,{{\bar {\cal E}}^2}\,
{{\partial {\omega_0}}\over {\partial {\cal B}}} ({\cal B}_0)
\,-\, \int\,{\bar {\cal M}} ({{\cal B}_0}\,,\,{\bar \Theta}\,,\,t)\,
d{\bar \Theta}\biggr]\>,
\eqno  (A10)
$$
\vskip 2pt
\noindent
one can show that equations (A7) and (A8) can be written as
\vskip  2pt
$$
{\dot{\bar{\cal E}}}\,=\,-\,{{\partial {\cal H}}\over {\partial {\bar \Theta}}}
\qquad\qquad,\qquad\qquad
{\dot{\bar \Theta}}\,=\,{{\partial {\cal H}}\over {\partial {\bar{\cal E}}}}\>.
\eqno  (A11)
$$
\vskip 3pt
\noindent
Equations (A11) imply that $\cal H$ can be considered 
as the Hamiltonian of the first-order partially averaged 
system at resonance. To this Hamiltonian, one can attribute
a potential function given by
\vskip  2pt
$$
V({\bar \Theta})\,=\,-\, 
\int\,{\bar {\cal M}} ({{\cal B}_0}\,,\,{\bar \Theta})\,d{\bar \Theta}\>.
\eqno (A12)
$$
\vskip  2pt
\noindent
Differentiating equations (A11) with respect to $t$ and using
the Hamiltonian $\cal H$, one can write
\vskip 1pt
$$
{\ddot {\bar\Theta}}\,-\,\varepsilon\,
\Bigl[{{\partial {\omega_0}}\over {\partial {\cal B}}} ({\cal B}_0)\Bigr]
\,{\bar {\cal M}}({{\cal B}_0},{\bar \Theta})\,=\,0\,.
\eqno (A13)
$$
\vskip 1pt
\noindent
Equation (A13) can be regarded as the equation of a mathematical
pendulum with Hamiltonian $\cal H$ and potential function
$V({\bar \Theta})$. 
The librational motion of this pendulum presents a geometrical interpretation 
for the resonance capture phenomenon. The maximum variation of the action
variable $\cal B$ associated
with these librational motions is given by (Wiggins 1996)
\vskip  2pt
$$
\Delta {\cal B}\,=\,2\,{\biggl\{2\,\varepsilon\,
{\Bigl[{{\partial {\omega_0}}\over {\partial {\cal B}}}({{\cal B}_0})\Bigr]^{-1}}
\,\Bigl[{V_{Max}}({\bar \Theta})\,-\,{V_{min}}({\bar \Theta})\Bigr]\biggr\}^{1/2}}
\,+\,O(\varepsilon)\>.
\eqno (A14)
$$
\vskip  30pt
\noindent
{\bigrmsixteen APPENDIX  B}
\vskip  10pt

From definition of $H$, we have
\vskip  2pt
$$
{{\partial H}\over {\partial L}}\,=\,
{1\over {|{\vec r}-{\bf{{\vec r}_1}}|^{-3}}}\,
\Biggl\{\Bigl[r- \cos (\theta-{\theta_1})\Bigr]\,
{{\partial r}\over {\partial L}}\,+\,
r \sin (\theta - {\theta_1})\,
{{\partial \theta}\over {\partial L}}\Biggr\}\,,
\eqno (B1)
$$
$$
{{\partial H}\over {\partial G}}\,=\,
{1\over {|{\vec r}-{\bf{{\vec r}_1}}|^{-3}}}\,
\Biggl\{\Bigl[r- \cos (\theta-{\theta_1})\Bigr]\,
{{\partial r}\over {\partial G}}\,+\,
r \sin (\theta - {\theta_1})\,
{{\partial \theta}\over {\partial G}}\Biggr\}\,,
\eqno (B2)
$$
$$\!\!\!\!\!
{{\partial H}\over {\partial \ell}}\,=\,
{1\over {|{\vec r}-{\bf{{\vec r}_1}}|^{-3}}}\,
\Biggl\{\Bigl[r- \cos (\theta-{\theta_1})\Bigr]\,
{{\partial r}\over {\partial \ell}}\,+\,
r \sin (\theta - {\theta_1})\,
{{\partial \theta}\over {\partial \ell}}\Biggr\}\,,
\eqno (B3)
$$
$$\!\!\!\!\!\!\!\!\!\!\!\!\!\!\!\!\!\!\!\!\!\!\!\!\!
\!\!\!\!\!\!\!\!\!\!\!\!\!\!\!\!\!\!\!\!\!\!\!\!\!
\!\!\!\!\!\!\!\!\!\!\!\!\!\!\!\!\!\!\!\!\!\!\!\!\!\!\!\!\!
{{\partial H}\over {\partial g}}\,=\,
{1\over {|{\vec r}-{\bf{{\vec r}_1}}|^{-3}}}\,
r \sin (\theta - {\theta_1})\,
{{\partial \theta}\over {\partial g}}\,.
\eqno (B4)
$$
\vskip  2pt
\noindent
From these equations it is evident that one needs to compute
derivatives of $r$ and $\theta$ with respect to all Delaunay
variables. From equation (3), we have
\vskip  2pt
$$\!\!\!\!\!\!\!\!\!\!\!\!\!\!\!\!\!\!\!\!\!\!\!\!\!
\!\!\!\!\!\!\!\!\!\!\!\!\!\!\!\!\!\!\!\!\!\!\!
{{\partial r}\over {\partial L}}\,=\,
-\,{\Bigl({G\over {1+e\cos v}}\Bigr)^2}\,
\Bigl(\cos v\,{{\partial e}\over {\partial L}}-\,
e \sin v\,{{\partial v}\over {\partial L}}\Bigr)
\eqno  (B5)
$$
$$
{{\partial r}\over {\partial G}}\,=\,
2\,\Bigl({G\over {1+e\cos v}}\Bigr)\,-\,
{\Bigl({G\over {1+e\cos v}}\Bigr)^2}\,
\Bigl(\cos v\,{{\partial e}\over {\partial G}}\,-\,
e\,\sin v\,{{\partial v}\over {\partial G}}\Bigr)\,,
\eqno  (B6)
$$
$$\!\!\!\!\!\!\!\!\!\!\!\!\!\!\!\!\!\!\!\!\!\!\!\!
\!\!\!\!\!\!\!\!\!\!\!\!\!\!\!\!\!\!\!\!\!\!\!\!
\!\!\!\!\!\!\!\!\!\!\!\!\!\!\!\!\!\!\!\!\!\!\!\!
\!\!\!\!\!\!\!\!\!\!\!\!\!\!\!\!\!\!\!\!\!\!\!\!
\!\!\!\!\!\!\!\!\!\!\!\!\!\!\!\!\!\!\!\!\!\!\!\!
\!\!\!\!\!\!\!\!
{{\partial r}\over {\partial \ell}}\,=\,e a \sin u\,
{{\partial u}\over {\partial \ell}}\,
\eqno  (B7)
$$
\vskip  2pt
\noindent
and $\partial r/\partial g =0$. On the other hand,
from $\theta = g+v$, $\partial \theta/\partial g = 1$ and
the  derivatives of $\theta$ with respect
to $L,G$ and $\ell$ will be equal to derivatives of $v$
with respect to these variables. 
Using $\ell=u-e\sin u$ and
$G=L{(1-{e^2})^{1/2}}$ along with equation (3), the partial
derivatives of $r$ with respect to the Delaunay variables
can be written as
\vskip  2pt
$$\!\!\!\!\!\!\!\!\!\!\!\!\!\!\!\!\!\!\!\!\!\!\!\!\!
{{\partial r}\over {\partial L}}\,=\,{r\over e}\,{a^{-1/2}}\,
\bigl(2e-\cos v -e{\cos^3} v\bigr)\,,
\eqno  (B8)
$$
$$\!\!\!\!\!\!\!\!\!\!\!\!\!\!\!\!\!\!\!\!\!\!\!\!\!
\!\!\!\!\!\!\!\!\!\!\!\!\!\!\!\!\!\!\!\!\!\!
{{\partial r}\over {\partial G}}\,=\,{1\over e}\,
{\bigl[a (1-{e^2})\bigr]^{1/2}}\,\cos v\>,
\eqno  (B9)
$$
$$\!\!\!\!\!\!\!\!\!\!\!\!\!\!\!\!\!\!\!\!\!\!\!\!\!
\!\!\!\!\!\!\!\!\!\!\!\!\!\!\!\!\!\!\!\!\!\!\!\!\!
{{\partial r}\over {\partial \ell}}\,=\,e\,a\,{(1-{e^2})^{-1/2}}\,\sin v\,,
\eqno  (B10)
$$
\vskip  2pt
\noindent
and the partial derivatives of $\theta$ with respect to $L,G$
and $\ell$ will be equal to
\vskip 2pt
$$\!\!\!\!\!\!\!\!\!\!\!\!\!\!\!\!\!\!\!\!\!\!\!\!\!\!
{{\partial \theta}\over {\partial L}}\,=\,
{{G^2}\over {e\,{L^3}\,(1-{e^2})}}\,\sin v\,(2+e\cos v)\,,
\eqno  (B11)
$$
$$
{{\partial \theta}\over {\partial G}}\,=\,-\,{1\over e}\, 
{\bigl[a(1-{e^2})\bigr]^{1/2}}\,\Bigl[{1\over r}\,+\,
{1\over {a(1-{e^2})}}\Bigr]\,\sin v\>,
\eqno  (B12)
$$
$$\!\!\!\!\!\!\!\!\!\!\!\!\!\!\!\!\!\!\!\!\!\!\!\!
\!\!\!\!\!\!\!\!\!\!\!\!\!\!\!\!\!\!\!\!\!\!\!\!
\!\!\!\!\!\!\!\!\!\!\!\!\!\!\!\!
{{\partial \theta}\over 
{\partial \ell}}\,=\,{\Bigl({a\over r}\Bigr)^2}\,{(1-{e^2})^{1/2}}\>.
\eqno  (B13)
$$
\vskip  2pt
\noindent
Replacing the derivatives of $r$ and $\theta$ in equations
(B1) to (B4) by their equivalent expressions given by
equations (B8) to (B13), one can write
\vskip 2pt
$$\!\!\!\!\!\!\!\!\!\!\!\!\!\!\!\!\!\!\!\!\!\!\!\!
\eqalign {
{{\partial H}\over {\partial L}}\,=\,{r\over e}\,{a^{-1/2}}\,
\Biggl\{&\biggl[{{r-\cos (\theta-{\theta_1})}\over
{|{\vec r}\,-\,{\bf{{\vec r}_1}}|^3}}\biggr]\,
\bigl(2e-\cos v -e{\cos^3} v\bigr)\cr
&\qquad\quad\qquad\qquad\quad\>\>
+\,\biggl[{{\sin (\theta - {\theta_1})}\over
{|{\vec r}\,-\,{\bf{{\vec r}_1}}|^3}}\biggr]\,
(2+e\cos v)\, \sin v\Biggr\}\>,\cr}
\eqno (B14)
$$
\vskip  6pt
$$\eqalign {
{{\partial H}\over {\partial G}}\,=\,{1\over e}\,{\bigl[a (1-{e^2})\bigr]^{1/2}}\,
\Biggl\{&\biggl[{{r\,-\,\cos ({\theta}-{\theta_1})}\over
{|{\vec r}\,-\,{\bf{{\vec r}_1}}|^3}}\biggr]\,\cos v\cr
&\qquad\qquad\>\>\>
-\,\biggl[{{\sin (\theta - {\theta_1})}\over
{|{\vec r}\,-\,{\bf{{\vec r}_1}}|^3}}\biggr]\,
\Bigl[{1\over r}\,+\,{1\over {a(1-{e^2})}}\Bigr]\,\sin v\Biggr\}\>,\cr}
\eqno  (B15)
$$
\vskip 6pt
$$
{{\partial H}\over {\partial \ell}}\,=\,a\,{(1-{e^2})^{-1/2}}\,
\Biggl\{e \sin v\,\biggl[{{r\,-\,\cos ({\theta}-{\theta_1})}\over
{|{\vec r}\,-\,{\bf{{\vec r}_1}}|^3}}\biggr]\,+\,
{a\over r}\,(1-{e^2})\,\biggl[{{\sin ({\theta}-{\theta_1})}\over
{|{\vec r}\,-\,{\bf{{\vec r}_1}}|^3}}\biggr]\Biggr\}\>,
\eqno (B16)
$$
\vskip  9pt
\noindent
and
\vskip  6pt
$$\!\!\!\!\!\!\!\!\!\!\!\!\!\!\!\!\!\!\!\!\!\!\!\!
\!\!\!\!\!\!\!\!\!\!\!\!\!\!\!\!\!\!\!\!\!\!\!\!
\!\!\!\!\!\!\!\!\!\!\!\!\!\!\!\!\!\!\!\!\!\!\!\!
\!\!\!\!\!\!\!\!\!\!\!\!\!\!\!\!\!\!\!\!\!\!\!\!
\!\!\!\!\!\!\!\!\!\!\!\!\!\!\!\!\!\!\!\!\!\!\!\!
\!\!\!\!\!\!\!\!\!\!\!\!\!\!\!\!\!\!\!\!\!\!\!\!
\!\!\!\!\!\!\!\!\!\!\!\!\!\!\!\!\!\!\!\!\!\!\!\!
{{\partial H}\over {\partial g}}\,=\,
{{r\,\sin (\theta - {\theta_1})}\over {|{\vec r}\,-\,{\bf{{\vec r}_1}}|^3}}\,,
\eqno  (B9)
$$
\vskip  16pt
\noindent
which along with equations (12) immediately result in expressions
(16) and (17).

\bye